\documentclass[aip,jmp,amsmath,amssymb,reprint]{revtex4-1}
\usepackage{braket,nicefrac}
\usepackage{graphicx,soul}
\usepackage{pgfpages}

\begin{document}
\title{Exact Orbital-Free Kinetic Energy Functional for General Many-Electron Systems}
\author{Thomas Pope}
\email{Thomas.Pope2@newcastle.ac.uk}
\affiliation{School of Natural and Environmental Sciences, Newcastle University, Newcastle NE1 7RU, United Kingdom}
\author{Werner Hofer}
\email{Werner.Hofer@newcastle.ac.uk}
\affiliation{School of Natural and Environmental Sciences, Newcastle University, Newcastle NE1 7RU, United Kingdom}
\affiliation{School of Physical Sciences, University of Chinese Academy of Sciences, Beijing 10090, China}
\date{\today}
\begin{abstract}
The exact form of the kinetic energy functional has remained elusive in orbital-free models of density functional theory (DFT). This has been the main stumbling block for the development of a general-purpose framework on this basis. Here, we show that on the basis of a two-density model, which represents many-electron systems by mass density and spin density components, we can derive the exact form of such a functional. The exact functional is shown to contain previously suggested functionals to some extent, with the notable exception of the Thomas-Fermi kinetic energy functional.
\end{abstract}
\maketitle

Orbital-free density functional theory has the potential to vastly increase the computational efficiency of electronic structure calculations \cite{levy1979universal,levy1984exact,pearson1993ab,wesolowski2013recent,lehtomaki2014orbital,karasiev2015frank}. However, a general kinetic energy functional has remained elusive \cite{garcia2008approach,huang2010nonlocal,shin2014enhanced,mi2018nonlocal,constantin2018semilocal,constantin2018nonlocal}.
It has even been shown that a perturbative approach to the construction of such a functional would diverge~\cite{seidl2000simulation}.
This has been seen, by more traditional DFT theorists, as an indication that the whole endeavour is ultimately doomed.

While approximations to the kinetic energy functionals exist for specific systems, a universal functional
does so far not exist. However, given the large amount of work and ingenuity spent on this research over the last
thirty years, it could be that the lack of decisive progress is due not to the intractability of nature at this scale,
 but due to restrictions in the very foundations on which the theoretical models have been constructed.

We have, in a number of publications  (see \cite{pope2017spin,pope2018two} and references therein), advocated a two-density model, where the general properties of a many-electron system are represented by mass density $\rho$ and spin density $S$ components. The spin density components in this case are associated with a chiral - and complex - vector field. These two components combined yield the conventional DFT density $n({\bf r})$:
\begin{equation}
n({\bf r}) = \rho({\bf r}) + S({\bf r}).
\end{equation}
It has been shown that such a two-density model also complies with theorems equal to the Hohenberg-Kohn theorems for a single density \cite{hohenberg1964inhomogeneous,pope2018two}. Moreover, it allowed us to define an {\it effective} wavefunction combining the roots of these two densities in the following way:
\begin{equation}
\Psi({\bf r})=\rho^{\nicefrac{1}{2}}({\bf r})+{\bf i}{\bf e}_S({\bf r})S^{\nicefrac{1}{2}}({\bf r}).
\end{equation}
Here, ${\bf e}_S$ a unit vector in the direction of the spin vector field and ${\bf i}{\bf e}_S$ is a unit bivector in the plane perpendicular to the spin vector. Such an effective wavefunction will be subject to an altered Hamiltonian, essentially due to the multivector properties of the effective wavefunction \cite{doran2003geometric}. The modified Hamiltonian is then given by (in atomic units):
\begin{equation}
H({\bf r})=-\frac{1}{2}\nabla^2+v_0({\bf r})+{\bf i}{\bf e}_b({\bf r})v_b({\bf r})=:\hat{h}_0({\bf r})+{\bf i}{\bf e}_b({\bf r})v_b({\bf r}).
\end{equation}
Here, ${\bf e}_b({\bf r})$ is a unit vector in the direction associated with the potential $v_b({\bf r})$, which we termed the {\it bivector} potential due to its geometric properties. The approach yields a very compact formulation for the many electron problem under the assumption that ${\bf e}_S({\bf r}) \approx {\bf e}_S$, in other words a constant vector. Then the vector ${\bf e}_b({\bf r}) = {\bf e}_b = {\bf e}_S$ and the general problem is determined by the matrix equation (we omit the explicit dependency on the position in the following):
\begin{equation}
\begin{pmatrix}\hat{h}_0&-v_b\\v_b&\hat{h}_0\end{pmatrix}
\begin{pmatrix}\rho^{\nicefrac{1}{2}}\\S^{\nicefrac{1}{2}}\end{pmatrix}
=\mu
\begin{pmatrix}\rho^{\nicefrac{1}{2}}\\S^{\nicefrac{1}{2}}\end{pmatrix},
\end{equation}
where $\mu$ is the chemical potential as in the standard formulation. It can be seen that the bivector potential governs the interaction between the two density components. In case it vanishes the formulation simplifies to the single electron case and a conventional Schr\"odinger equation for single electrons. The bivector potential can be explicitly
derived from the gradients of the two density components. It is \cite{pope2018two}:
\begin{equation}
v_b=\frac{1}{\rho+S}\left[S^{\nicefrac{1}{2}}\left(-\frac{\nabla^2}{2}\right)\rho^{\nicefrac{1}{2}}-\rho^{\nicefrac{1}{2}}\left(-\frac{\nabla^2}{2}\right) S^{\nicefrac{1}{2}}\right].
\end{equation}
However, it should be noted that the theoretical framework described by the last two equations is a simplification, as it assumes the unit vector of the spin density fields to be constant. In the following, we shall describe model simulations with this simplified model. The rest of the paper is then devoted to removing the simplification. The exact formulation of the problem, as will be shown, also leads to an exact kinetic energy functional.

We have implemented the simplified model in the self-consistency cycle of the DFT package CASTEP \cite{clark2005first}. Similar to the auxiliary functional approach \cite{hasnip2015auxiliary}, the model takes the Kohn-Sham \cite{kohn1965self} density at each stage of the self-consistency cycle, independently minimises it according to the two-density many-body equations and returns a density to the parent Kohn-Sham cycle. Convergence occurs when consecutive densities returned by the two-density model are equivalent and thus the results of the two models are consistent. Since correlation is not yet present in our model, we employ a Local-Density-Approximation (LDA) correlation term \cite{perdew1981self}.

\begin{figure}
  \centering
  \includegraphics[width=\linewidth]{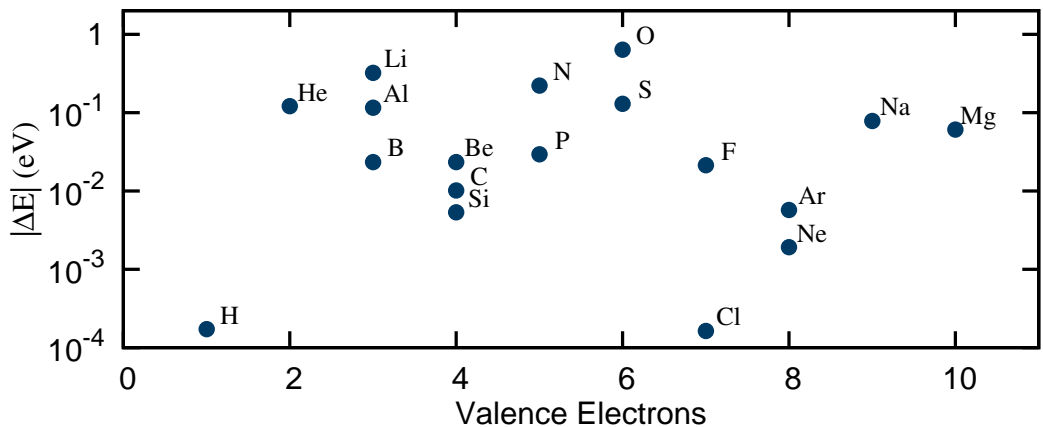}
  \caption{Comparison of total energy simulations of eighteen atomic systems with conventional Kohn-Sham DFT methods and the two-density model. The energy differences are in all cases very small. Note that we compare total energies including all electrons of the system. Data adapted from Pope \emph{et al.}~\cite{pope2018two} with permission from Springer Nature, copyright 2019.}\label{figure1}
\end{figure}

Figure \ref{figure1} shows the result of model simulations for eighteen atomic systems. The figure shows the absolute values of the energy differences obtained with standard Kohn-Sham DFT methods and with the two-density model. The differences are in all cases very small. We can thus say that the simplified model already agrees with the total energy values obtained with standard DFT methods. The agreement is very good not only for the atomic systems shown here, but also for the interatomic distances in dimers, and for the bandstructure of semiconductors which is very sensitive to variations of the electron density. These simulations are shown elsewhere \cite{pope2018two}.

It should be noted that the two-density approach no longer contains exchange functionals as in standard DFT. Within our approach, exchange energy is an emergent property, which is due to the interaction between the two densities. So it seems that two very different approaches to a general many-electron system lead to the same physical properties. Moreover, the agreement, in case it persists for more complicated systems, raises questions about the level of approximation present in standard DFT methods. Because within the two-density model the results were derived under the assumption that the vector fields for both, spin-densities and bivector potentials, are independent of location and of constant direction. In general, this will not be the case.

It is difficult, at this point in the development of the two-density method, to gauge the importance of this finding. Generally, it had been assumed within the DFT community, that only the exchange-correlation functionals are subject to approximations. If we knew their exact form, then we would have an in principle exact method of simulating many-electron systems, was the general opinion so far. This result, however, suggests that there is another level of approximation involved, which does not depend on the exchange-correlation functionals at all. And this level concerns the change of local physical properties. If the direction of the vector associated with the spin density, which is one of the two density components, changes faster than a certain threshold, then standard DFT methods will no longer be sufficient to describe the dynamics of the system. In the following we shall derive the exact functional and show the approximations involved in the two-density scalar kinetic energy, which is equivalent to the standard DFT formalism.

In general both vectors, ${\bf e}_S$ and ${\bf e}_b$ will depend on the position so that ${\bf e}_S = {\bf e}_S({\bf r})$ and ${\bf e}_b = {\bf e}_b({\bf r})$. This means that both, the roots of the spin density and the bivector potential will consist of three individual components. In a cartesian basis we may write the term associated with the spin density as:
\begin{equation}
{\bf i}{\bf e}_S({\bf r})S^{\nicefrac{1}{2}}({\bf r})=\sum_{j=1}^3{\bf i}{\bf e}_j S_j^{\nicefrac{1}{2}}({\bf r}).
\end{equation}
Correspondingly, the term with the bivector potential will, in the general case, be:
\begin{equation}
{\bf i}{\bf e}_b({\bf r})v_b({\bf r})=\sum_{j=1}^3{\bf i}{\bf e}_j v_j({\bf r}).
\end{equation}
The general formulation is quite complicated, and the detailed analytical calculation is omitted for brevity. It can be found in the supplemental material at \S~\ref{bivec_pot}. The ansatz leads to a set of four differential equations. The equations in matrix notation are given by:
\begin{equation}
 \begin{pmatrix}
  \hat{h}_0&-v_1&-v_2&-v_3\\
  v_1&\hat{h}_0&v_3&-v_2\\
  v_2&-v_3&\hat{h}_0&v_1\\
  v_3&v_2&-v_1&\hat{h}_0
 \end{pmatrix}\!\!\!
 \begin{pmatrix}
  \rho^{\nicefrac{1}{2}}\\S_1^{\nicefrac{1}{2}}\\S_2^{\nicefrac{1}{2}}\\S_3^{\nicefrac{1}{2}}
 \end{pmatrix}
 =\mu
 \begin{pmatrix}
  \rho^{\nicefrac{1}{2}}\\S_1^{\nicefrac{1}{2}}\\S_2^{\nicefrac{1}{2}}\\S_3^{\nicefrac{1}{2}}
 \end{pmatrix}.
\end{equation}
The components $v_k$ of the bivector potential are described by the following expression:
\begin{equation}
 v_k=\frac{1}{4} \frac{\hat{\bf J}_k-B_k^j\hat{\bf J}_i-B_k^i\hat{\bf J}_j}{1-B_k^j\chi_i^+-B_k^i\chi_j^-},
\end{equation}
where the differential operator $\hat{\bf J}_k$ is given by the expression:
\begin{align}
 \hat{\bf J}_k&=\left(\frac{\rho-S}{2}+S_k\right)^{-1}\times\nonumber\\&\text{\hspace{1cm}}
 \nabla \left(\rho^{\nicefrac{1}{2}}\nabla S_k^{\nicefrac{1}{2}}-S_k^{\nicefrac{1}{2}}\nabla\rho^{\nicefrac{1}{2}}+S_i^{\nicefrac{1}{2}}\nabla S_j^{\nicefrac{1}{2}}-S_j^{\nicefrac{1}{2}}\nabla S_j^{\nicefrac{1}{2}}\right)
\end{align}
and the mixing parameters are given by,
\begin{align}
 B_k^j&=\frac{\chi_j^--\chi_i^+\chi_k^+}{1-\chi_k^-\chi_k^+},\qquad
 B_k^i=\frac{\chi_i^+-\chi_j^-\chi_k^-}{1-\chi_k^-\chi_k^+}\nonumber\\
 \chi_k^{\pm}&=\left(S_i^{\nicefrac{1}{2}}S_j^{\nicefrac{1}{2}}\pm\rho^{\nicefrac{1}{2}}S_k^{\nicefrac{1}{2}}\right) \left(\frac{\rho-S}{2}+S_k\right)^{-1}
\end{align}
Here, the indices are defined by the bivector notation, ${\bf i}{\bf e}_k={\bf e}_i{\bf e}_j$.

The formulation leads to a very straightforward evaluation of individual energy components of the system.
The total energy density is given by,
\begin{equation}
  \epsilon_{\text{tot}}=\rho^{\nicefrac{1}{2}}\hat{h}_0\rho^{\nicefrac{1}{2}}+\sum_kS_k^{\nicefrac{1}{2}}\hat{h}_0S_k^{\nicefrac{1}{2}}\label{KS4},
\end{equation}
where the same Hamiltonian operator is used on each of the four density components (details may be found in \S~{\ref{app_kin}}). This operator consists of a kinetic energy operator and a potential associated with the electron-electron and electron-nuclear Coulomb interactions. This leads to three constituent energy densities, the electron-nuclear energy density,
\begin{equation}
 \varepsilon_{\text{e-n}}\!\left({\bf r}\right)=v_{\text{e-n}}\!\left({\bf r}\right)n\!\left({\bf r}\right),
\end{equation}
the electron-electron energy density,
\begin{equation}
 \varepsilon_{\text{e-e}}\!\left({\bf r}\right)=v_{\text{e-e}}\!\left({\bf r}\right)n\!\left({\bf r}\right),
\end{equation}
and the kinetic energy density,
\begin{equation}
\varepsilon_{\text{k}}({\bf r})=-\frac{1}{4}\nabla^2 n({\bf r}) + \frac{1}{8} \left(
 \frac{\left[\nabla\rho({\bf r})\right]^2}{\rho({\bf r})}+
 \sum_{k=1}^3 \frac{\left[\nabla S_k({\bf r})\right]^2}{S_k({\bf r})}
 \right).
\end{equation}
This form of the energy density has some similarities to previous forumlations, which become clearer when the equation is reformulated. We define the kinetic energy density as a sum of three distinct terms,
\begin{equation}
 \varepsilon_{\text{k}}({\bf r})=\varepsilon_{\text{k}}^1({\bf r})+\varepsilon_{\text{k}}^2({\bf r})+\varepsilon_{\text{k}}^3({\bf r}).
\end{equation}
Here the first term requires only the total charge density, $n$, and no further information,
\begin{equation}
 \varepsilon_{\text{k}}^1({\bf r})=-\frac{1}{4}\nabla^2 n({\bf r})+\frac{1}{8}\frac{\left(\nabla n({\bf r})\right)^2}{n({\bf r})}.
\end{equation}
This is equivalent to the single density formulation presented by Levi \emph{et al} (LPS)~\cite{levy1984exact}.

The second term contains the mass and spin density scalar quantities, $\rho$ and $S$,
\begin{equation}
 \varepsilon_{\text{k}}^2({\bf r})=\frac{1}{8}\left(\frac{\left[\nabla(\rho({\bf r})/n({\bf r}))\right]^2}{\rho({\bf r})/n({\bf r})}+\frac{\left[\nabla(S({\bf r})/n({\bf r}))\right]^2}{S({\bf r})/n({\bf r})}
 \right)n({\bf r}).
\end{equation}
This is equivalent to the kinetic energy density correction presented in the scalar two-density model (2D)~\cite{pope2018two}.

The final term includes the vector information for the spin density,
\begin{align}
 \varepsilon_{\text{k}}^3\left({\bf r}\right)&=\frac{1}{8}\sum_{k=1}^3\frac{\left[\nabla(S_k({\bf r})/S({\bf r}))\right]^2}{S_k({\bf r})/S({\bf r})}\nonumber\\&\text{\hspace{2cm}}
 +\frac{1}{4}\sum_{k=1}^3\nabla(S_k({\bf r})/S({\bf r}))\nabla S({\bf r}).
\end{align}
This final term completes the kinetic energy density and reveals the approximations that have been made by the previous approaches. Namely, the LPS and 2D approaches can be written respectively,
\begin{align}
 \varepsilon_{\text{k}}^{\text{LPS}}({\bf r})&=\varepsilon_{\text{k}}^1({\bf r}),\nonumber\\
 \varepsilon_{\text{k}}^{\text{2D}}({\bf r})&=\varepsilon_{\text{k}}^1({\bf r})+\varepsilon_{\text{k}}^2({\bf r}).
\end{align}

A functional for each energy component is given by integrating the energy densities,
\begin{align}
 E_{\text{e-n}}&=\int{v_{\text{e-n}}\!\left({\bf r}\right)n\!\left({\bf r}\right)\text{d}{\bf r}},\nonumber\\
 E_{\text{e-e}}&=\int{v_{\text{e-e}}\!\left({\bf r}\right)n\!\left({\bf r}\right)\text{d}{\bf r}},\nonumber\\
 E_{\text{kin}}&=-\frac{1}{4}\int{\nabla^2 n({\bf r})\text{d}{\bf r}}\nonumber\\
 &\text{\hspace{1cm}}+\frac{1}{8}\int{\left(
 \frac{\left[\nabla\rho({\bf r})\right]^2}{\rho({\bf r})}+
 \sum_{k=1}^3 \frac{\left[\nabla S_k({\bf r})\right]^2}{S_k({\bf r})}
 \right)
 \text{d}{\bf r}}.
\end{align}

It is noteworthy that the only energy term, which contains the individual components of the total density, is the kinetic energy. Moreover, it contains these components in a form which is very familiar in orbital-free
DFT. The original von Weizs\"acker correction to the kinetic energy functional has the form \cite{von1935theory}:

\begin{equation}
T_{vw}=\frac{1}{8}\int\frac{(\nabla n)^2}{n},
\end{equation}
where $n$ is the electron (mass or charge) density. It can be seen that the two-density model leads to similar terms. This suggests that the correction has a more general significance than previously assumed. Here, it is not just a term introduced from a single-electron case, that is the solution of the hydrogen problem, but a term which is always present in a many-electron system. This derivation also indicates that it would be largely impossible to capture this behaviour of a many-electron system with a single density only, as has been attempted in the past. An estimate of the exact functional also shows that its kernel is at most linear with the density. The functional for free electrons, by contrast, would involve the DFT density to the power of 5/3. From this perspective it is quite clear that a model starting from free electrons will lead to instabilities as soon as the density increases beyond a certain level.

We note that this kinetic energy functional is the kinetic energy functional of interacting electrons, not non-interacting electrons as in the Kohn-Sham formalism. Furthermore, it is easy to establish that the two-density framework extends into the regime of single free electrons, in contrast to single-density DFT, where the kinetic energy of free electrons cannot be derived from the density. Here, the density roots of free electrons are (we set the spin vector equal to the $x$-direction):
\begin{equation}
\rho^{\nicefrac{1}{2}}(x)=A\sin{kx},\qquad S_1^{\nicefrac{1}{2}}(x)=S^{\nicefrac{1}{2}}(x)=A\cos{kx}.
\end{equation}
Then the components of the kinetic energy density read:
\begin{align}
\varepsilon_{\text{k}}^1({\bf r})&=0,\qquad
\varepsilon_{\text{k}}^2({\bf r})=\frac{1}{2}A^2k^2,\qquad
\varepsilon_{\text{k}}^3({\bf r})=0.
\end{align}
and the kinetic energy functional yields $E_{\text{kin}}=k^2/2$, because $A^2=1/V$. Notable, the energy derived from a single-density perspective, $\varepsilon_{\text{k}}^{1}$, is zero. From a physical perspective this means that there is continuity in the new framework
going from single non-interacting electrons to a more complicated system where electrons interact. The
change from the former to the latter can then be as incremental as one would wish. This should make the analysis of events at the local level much simpler in the future.

\section*{Acknowledgements}
The authors acknowledge EPSRC funding for the UKCP consortium (Grant No. EP/K013610/1). This work
was also supported by the North East Centre for Energy Materials (NECEM). WH acknowledges support
from the University of Chinese Academy of Sciences.

\bibliography{ref}{}

\begin{widetext}
 \begin{appendix}
 \section{Supporting Material}
 \subsection{Bivector Potential}\label{bivec_pot}
 The Schr\"{o}dinger equation in the extended electron model is given,
 \begin{align}
  H\Psi
  &=\left(-\frac{1}{2}\nabla^2+v_0+\sum_l{\bf ie}_lv_l\right)\left(\rho^{\nicefrac{1}{2}}+\sum_k{\bf ie}_kS_k^{\nicefrac{1}{2}}\right)\nonumber\\
  &=\underbrace{\left(-\frac{1}{2}\nabla^2\right)\rho^{\nicefrac{1}{2}}+v_0\rho^{\nicefrac{1}{2}}-\sum_lv_lS_l^{\nicefrac{1}{2}}}_{s_0}+
  \sum_k{\bf ie}_k\underbrace{\left[\left(-\frac{1}{2}\nabla^2\right)S_k^{\nicefrac{1}{2}}+v_0S_k^{\nicefrac{1}{2}}+v_k\rho^{\nicefrac{1}{2}}-\left(v_iS_j^{\nicefrac{1}{2}}-v_jS_i^{\nicefrac{1}{2}}\right)\right]}_{b_k}.
 \end{align}
 Here, the indices $i$ and $j$ are defined by bivector notation, such that ${\bf ie}_k={\bf e}_i{\bf e}_j$. This yields four equations, the scalar and the three bivector parts, which in matrix form are given,
\begin{equation}
 \begin{pmatrix}
  \hat{h}_0&-v_1&-v_2&-v_3\\v_1&\hat{h}_0&v_3&-v_2\\v_2&-v_3&\hat{h}_0&v_1\\v_3&v_2&-v_1&\hat{h}_0
 \end{pmatrix}
 \begin{pmatrix}
  \rho^{\nicefrac{1}{2}}\\ S_1^{\nicefrac{1}{2}}\\ S_2^{\nicefrac{1}{2}}\\ S_3^{\nicefrac{1}{2}}
 \end{pmatrix}
 =\mu
 \begin{pmatrix}
  \rho^{\nicefrac{1}{2}}\\ S_1^{\nicefrac{1}{2}}\\ S_2^{\nicefrac{1}{2}}\\ S_3^{\nicefrac{1}{2}}
 \end{pmatrix}.
\end{equation}

Multiplying by the dual, we find,
\begin{equation}
 \Psi^{\dagger}H\Psi
 =\left(\rho^{\nicefrac{1}{2}}-\sum_l{\bf ie}_lS_l^{\nicefrac{1}{2}}\right)\left(s_0+\sum_k{\bf ie}_kb_k\right)
 =\rho^{\nicefrac{1}{2}}s_0+\sum_lS_l^{\nicefrac{1}{2}}b_l+
 \sum_k{\bf ie}_k\left[
 -S_k^{\nicefrac{1}{2}}s_0+\rho^{\nicefrac{1}{2}}b_k+\left(S_i^{\nicefrac{1}{2}}b_j-S_j^{\nicefrac{1}{2}}b_i\right)
 \right].
\end{equation}
The scaler part may be simplified,
\begin{equation}
 \rho^{\nicefrac{1}{2}}s_0+\sum_kS_k^{\nicefrac{1}{2}}b_k
 =\rho^{\nicefrac{1}{2}}\left(-\frac{1}{2}\nabla^2\right)\rho^{\nicefrac{1}{2}}+
 \sum_kS_k^{\nicefrac{1}{2}}\left(-\frac{1}{2}\nabla^2\right)S_k^{\nicefrac{1}{2}}+v_0\left(\rho+\sum_kS_k\right)-\sum_kS_k^{\nicefrac{1}{2}}\left(v_iS_j^{\nicefrac{1}{2}}-v_jS_i^{\nicefrac{1}{2}}\right).
\end{equation}
We note that one can show,
\begin{equation}
 \sum_kS_k^{\nicefrac{1}{2}}\left(v_iS_j^{\nicefrac{1}{2}}-v_jS_i^{\nicefrac{1}{2}}\right)=0.
\end{equation}
The bivector part may also be simplified,
\begin{align}
 -S_k^{\nicefrac{1}{2}}s_0+\rho^{\nicefrac{1}{2}}b_k
 &=\rho^{\nicefrac{1}{2}}\left(-\frac{1}{2}\nabla^2\right)S_k^{\nicefrac{1}{2}}-S_k^{\nicefrac{1}{2}}\left(-\frac{1}{2}\nabla^2\right)\rho^{\nicefrac{1}{2}}+\sum_lv_lS_l^{\nicefrac{1}{2}}S_k^{\nicefrac{1}{2}}+v_k\rho-\rho^{\nicefrac{1}{2}}\left(v_iS_j^{\nicefrac{1}{2}}-v_jS_i^{\nicefrac{1}{2}}\right)
 \nonumber\\
 S_i^{\nicefrac{1}{2}}b_j-S_j^{\nicefrac{1}{2}}b_i
 &=S_i^{\nicefrac{1}{2}}\left(-\frac{1}{2}\nabla^2\right)S_j^{\nicefrac{1}{2}}-S_j^{\nicefrac{1}{2}}\left(-\frac{1}{2}\nabla^2\right)S_i^{\nicefrac{1}{2}}
 -\rho^{\nicefrac{1}{2}}\left(v_iS_j^{\nicefrac{1}{2}}-v_jS_i^{\nicefrac{1}{2}}\right)
 -v_k\left(S_i+S_j\right)+\sum_{l\neq k}v_lS_l^{\nicefrac{1}{2}}S_k^{\nicefrac{1}{2}}
\end{align}
Here we introduce the notation,
\begin{equation}
 \hat{\bf J}_k=\left(\frac{\rho-S}{2}+S_k\right)^{-1}\nabla \left(\rho^{\nicefrac{1}{2}}\nabla S_k^{\nicefrac{1}{2}}-S_k^{\nicefrac{1}{2}}\nabla\rho^{\nicefrac{1}{2}}+S_i^{\nicefrac{1}{2}}\nabla S_j^{\nicefrac{1}{2}}-S_j^{\nicefrac{1}{2}}\nabla S_j^{\nicefrac{1}{2}}\right)
\end{equation}
So the bivector term can be written,
\begin{equation}
 -S_k^{\nicefrac{1}{2}}s_0+\rho^{\nicefrac{1}{2}}b_k+S_i^{\nicefrac{1}{2}}b_j-S_j^{\nicefrac{1}{2}}b_i
 =-\frac{1}{2}\left(\frac{\rho-S}{2}+S_k\right)\hat{\bf J}_k+v_k\left(\rho+S_k-S_i-S_j\right)-2\rho^{\nicefrac{1}{2}}\left(v_iS_j^{\nicefrac{1}{2}}-v_jS_i^{\nicefrac{1}{2}}\right)+2\sum_{l\neq k}v_lS_l^{\nicefrac{1}{2}}S_k^{\nicefrac{1}{2}}
\end{equation}
Since we know the total energy must be a scalar object, the bivector terms must be zero,
\begin{align}
 \frac{1}{4}\left(\frac{\rho-S}{2}+S_k\right)\hat{\bf J}_k
 &=v_k\left(\frac{\rho-S}{2}+S_k\right)
 +v_i\left(S_k^{\nicefrac{1}{2}}S_i^{\nicefrac{1}{2}}-\rho^{\nicefrac{1}{2}}S_j^{\nicefrac{1}{2}}\right)
 +v_j\left(S_j^{\nicefrac{1}{2}}S_k^{\nicefrac{1}{2}}+\rho^{\nicefrac{1}{2}}S_i^{\nicefrac{1}{2}}\right)\nonumber\\
 \frac{1}{4}\hat{\bf J}_k
 &=v_k+\chi_i^+v_j+\chi_j^-v_i,
\end{align}
where, we define,
\begin{equation}
 \chi_k^{\pm}=\left(S_i^{\nicefrac{1}{2}}S_j^{\nicefrac{1}{2}}\pm\rho^{\nicefrac{1}{2}}S_k^{\nicefrac{1}{2}}\right) \left(\frac{\rho-S}{2}+S_k\right)^{-1}
\end{equation}
so that,
\begin{equation}
 v_k=\frac{1}{4}\hat{\bf J}_k-\chi_i^+v_j-\chi_j^-v_i
\end{equation}
substituting the expresion for $v_j(v_i)$ into the equation for $v_i(v_j)$, we find
\begin{align}
 \left(1-\chi_k^-\chi_k^+\right)v_i&=\frac{1}{4}\hat{\bf J}_i-
 \frac{1}{4}\chi_k^-\hat{\bf J}_j+\left(\chi_k^-\chi_i^--\chi_j^+\right)v_k
 \nonumber\\
 \left(1-\chi_k^-\chi_k^+\right)v_j&=\frac{1}{4}\hat{\bf J}_j-\frac{1}{4}\chi_k^+\hat{\bf J}_i+\left(\chi_k^+\chi_j^+-\chi_i^-\right)v_k
\end{align}
and subtituting these expressions into the equation for $v_k$ yeilds,
\begin{equation}
 v_k=\frac{1}{4} \frac{\hat{\bf J}_k-B_k^j\hat{\bf J}_i-B_k^i\hat{\bf J}_j}{1-B_k^j\chi_i^+-B_k^i\chi_j^-},
\end{equation}
where
\begin{equation}
 B_k^j=\frac{\chi_j^--\chi_i^+\chi_k^+}{1-\chi_k^-\chi_k^+},\qquad
 B_k^i=\frac{\chi_i^+-\chi_j^-\chi_k^-}{1-\chi_k^-\chi_k^+}
\end{equation}

\subsection{Kinetic Energy Density}\label{app_kin}
The total energy density is given by Eq.{~\ref{KS4}},
\begin{equation}
  \epsilon_{\text{tot}}
  =\rho^{\nicefrac{1}{2}}\hat{h}_0\rho^{\nicefrac{1}{2}}+\sum_kS_k^{\nicefrac{1}{2}}\hat{h}_0S_k^{\nicefrac{1}{2}}
  =\rho^{-\nicefrac{1}{2}}\left(\frac{\nabla^2}{2}\right)\rho^{-\nicefrac{1}{2}}+\sum_kS_k^{\nicefrac{1}{2}}\left(\frac{\nabla^2}{2}\right)S_k^{\nicefrac{1}{2}}+v_0n({\bf r}),
\end{equation}
where $v_0$ contains the electron-electron and electron-nuclear potentials. The kinetic energy density is thus given,
\begin{equation}
 \epsilon_{\text{kin}}
 =\rho^{\nicefrac{1}{2}}\left(-\frac{\nabla^2}{2}\right)\rho^{\nicefrac{1}{2}}+\sum_kS_k^{\nicefrac{1}{2}}\left(-\frac{\nabla^2}{2}\right)S_k^{\nicefrac{1}{2}}
 =-\frac{1}{4}\nabla^2 n({\bf r}) + \frac{1}{8} \left(
 \frac{\left[\nabla\rho({\bf r})\right]^2}{\rho({\bf r})}+
 \sum_{k=1}^3 \frac{\left[\nabla S_k({\bf r})\right]^2}{S_k({\bf r})}
 \right)
\end{equation}

Firstly, let $S_k=\varphi_kS$,
\begin{equation}
  \epsilon_{\text{kin}}
  =-\frac{1}{4}\nabla^2 n({\bf r}) + \frac{1}{8} \left(
 \frac{\left[\nabla\rho({\bf r})\right]^2}{\rho({\bf r})}+\frac{\left[\nabla S({\bf r})\right]^2}{S({\bf r})}\right)+\frac{1}{8}
 \sum_{k=1}^3\left(\frac{\left[\nabla\varphi_k({\bf r})\right]^2}{\varphi_k({\bf r})}+2\nabla\varphi_k({\bf r})\nabla S({\bf r})\right)
 \end{equation}
 Next, let $\rho=\chi n$ and $S=\left(1-\chi\right)n$,
\begin{align}
  \epsilon_{\text{kin}}&=
  -\frac{1}{4}\nabla^2 n({\bf r})
  +\frac{1}{8}\frac{\left(\nabla n({\bf r})\right)^2}{n({\bf r})}
  +\frac{1}{8}\frac{n({\bf r})}{1-\chi({\bf r})} \frac{\left(\nabla\chi({\bf r})\right)^2}{\chi({\bf r})}
  +\frac{1}{8}\sum_{k=1}^3\left(\frac{\left[\nabla\varphi_k({\bf r})\right]^2}{\varphi_k({\bf r})}+2\nabla\varphi_k({\bf r})\nabla \left(1-\chi({\bf r})\right)n({\bf r})\right),\nonumber\\
  &=\varepsilon_{\text{k}}^1\left({\bf r}\right)+\varepsilon_{\text{k}}^2\left({\bf r}\right)+\varepsilon_{\text{k}}^3\left({\bf r}\right).
 \end{align}
 where
 \begin{align}
  \varepsilon_{\text{k}}^1\left({\bf r}\right)&=-\frac{1}{4}\nabla^2 n({\bf r})
  +\frac{1}{8}\frac{\left(\nabla n({\bf r})\right)^2}{n({\bf r})},\nonumber\\
  \varepsilon_{\text{k}}^2\left({\bf r}\right)&=\frac{1}{8} \frac{n({\bf r})}{S({\bf r})} \frac{n({\bf r})}{\rho({\bf r})} \frac{\left(\nabla\chi({\bf r})\right)^2}{\chi({\bf r})}n({\bf r}),\nonumber\\
  \varepsilon_{\text{k}}^3\left({\bf r}\right)&=\frac{1}{8}\sum_{k=1}^3\left(\frac{\left(\nabla\frac{S_k({\bf r})}{S({\bf r})}\right)^2}{\left(\frac{S_k({\bf r})}{S({\bf r})}\right)}+2\left(\nabla\frac{S_k({\bf r})}{S({\bf r})}\right)\nabla S({\bf r})\right)
 \end{align}
\end{appendix}
\end{widetext}

\end{document}